\documentclass[11pt]{article}

\usepackage{epsfig}
\usepackage{graphicx}

\begin{document}

\baselineskip=16pt


\newcommand{\be}{\begin{equation}}
\newcommand{\ee}{\end{equation}}
\newcommand{\ba}{\begin{eqnarray}}
\newcommand{\ea}{\end{eqnarray}}
\newcommand{\no}{\nonumber \\}

\makeatletter
\renewcommand{\theequation}{\thesection.\arabic{equation}}
\@addtoreset{equation}{section}
\makeatother
\begin{titlepage}
\title{
\hfill\parbox{4cm}
{\normalsize {\tt hep-th/0610113}}\\
\vspace{1.5cm}
 {\bf  Elliptic Flow, Kasner Universe and Holographic Dual of RHIC Fireball}
}

\author{
Sang-Jin Sin$\,^{a,b}$
\thanks{{\tt E-mail: sjsin@hanyang.ac.kr}}
,
Shin Nakamura$\,^{a,c}$
\thanks{{\tt E-mail: nakamura@hanyang.ac.kr}}
$\ $and$\ $
Sang Pyo Kim$\,^{b,d}$
\thanks{{\tt E-mail: sangkim@kunsan.ac.kr}}
\vspace{5mm}\\[15pt]
$^a$~{\it BK21 Program Division in Physics, Department of physics,}\\{\it Hanyang University, Seoul 133-791, Korea}\\
$^b$~{\it Asia Pacific Center for Theoretical Physics, Postech, Pohang 790-784, Korea}\\
$^c$~{\it Center for Quantum Spacetime, Sogang University, Seoul 121-742, Korea}\\
$^d$~{\it Department of Physics, Kunsan National University, Kunsan 573-701, Korea}\\
\\[10pt]
}
\date{October 10, 2006}
\maketitle
\thispagestyle{empty}

\begin{abstract}
\normalsize \noindent
We consider a holographic dual of
hydrodynamics of ${\cal N}=4$ SYM plasma that undergoes
non-isotropic three-dimensional expansion relevant to RHIC fireball.
Our model is a natural extension of the Bjorken's one-dimensional
expansion, and it describes an elliptic flow whose $v_{2}$ and
eccentricity are given in terms of anisotropy parameters.
Holographic renormalization shows that absence of conformal anomaly
in the SYM theory constrains our local rest frame to be a Kasner
spacetime.
We show that the Kasner spcetime provides a simple
description of the anisotropically expanding fluid within a
well-controled approximation.
We also find that the dual geometry determines some of the hydrodynamic
quantities in terms of the initial condition and the fundamental
constants.

\end{abstract}
\end{titlepage}

\section{Introduction}
There has been much hope that one might be able to use  AdS/CFT
\cite{AdS/CFT} to describe the real systems after certain amount of
deformations. In fact it has been suggested that the fireball in
Relativistic Heavy Ion Collision (RHIC)
can be explained from dual gravity point of view
\cite{SZ,Nastase,SSZ,Aharony}, since the quark-gluon plasma (QGP)
created there are in the strong coupling region \cite{RHIC-1,RHIC-2}.
Although the YM theory described by the standard AdS/CFT
is large-$N_{c}$ ${\cal N}=4$ SYM theory,
there are many attempts to construct models closer to QCD
\cite{AdS/QCD}.
SUSY is not very relevant in the finite temperature context
because it is broken completely.

Since the RHIC fireball is expanding, we need to understand AdS/CFT
in time dependent situations. Recently, Janik and Peschanski
\cite{Janik, Janik-2} discussed this problem in non-viscous cases.
They use the conservation law and conformal invariance together with
the holographic renormalization \cite{HSS,Skend} to express the bulk
geometry from the given boundary data. As a result, the bulk
geometry reproduces the basic features of Bjorken theory
\cite{Bjorken}. Theses results were generalized  to the cases where
shear viscosity is included \cite{sin-shin}. Indeed, it had been
pointed out that inclusion of shear viscosity is very important in
the analysis of real RHIC physics since it plays an essential role
in the elliptic flow (see for example,
\cite{KH,bulk-zero,Teaney,Kolb,Hirano-Heinz,Zhang}). The shear
viscosity at the strong coupling limit was calculated for the ${\cal
N}=4$ SYM systems in \cite{PSS-1} using AdS/CFT, which fits the
perfect fluidity of RHIC QGP (see \cite{RHIC-2,bulk-zero,Hirano} and
the references therein).

In this paper, we consider a holographic dual of strongly interacting
${\cal N}=4$ large-$N_{c}$ SYM fluid with non-isotropic
three-dimensional expansion which is relevant to
``Little Bang'' of RHIC.
We will first make a simplest generalization of the Bjorken's
one-dimensional expansion to three-dimensional cases.
The resulting local rest frame of the fluid is described by
the Kasner metric
whose extreme limit reproduces the Bjorken hydrodynamics.
Interestingly, the hydrodynamic equation and the
equation of state will be shown to be independent of
the non-isotropy parameters.

Further more, using holographic
renormalization, we will establish the holographic dual
geometry of the anisotropically expanding fluid in the Kasner
spacetime and discuss its physical consequences by extending
the work of \cite{Janik,sin-shin}.
Interestingly, the gravity dual carries more information than
the hydrodynamics which is our input.
The integration constants coming from
the hydrodynamic equation cannot be determined by the hydrodynamics
itself. However, we will show that such quantities can be determined
by considering the dual geometry. Therefore it provides a method to
extract useful information on the macroscopic properties of the
expanding fluid in terms of microscopic data.

Kasner spacetime is a curved spacetime in general.
However, we find that the Kasner spacetime
can be a well-controled approximation of a local rest frame
of an anisotropically expanding fluid on Minkowski spacetime.
Therefore, the Kasner spacetime provides a useful framework
for analysing the three-dimensional expansion because of
the simplification of the hydrodynamic equations.

The organization of the paper is the following. In Section
\ref{hydro},
we introduce Kasner spacetime as a local rest frame of a fluid
under three-dimensional anisotropic expansion.
We also establish the hydrodynamics based on that local rest
frame. Section \ref{gravity} gives analyses in the gravity
dual in the late time regime.
We extend the results of \cite{Janik,sin-shin} to the
case of our interest: the three-dimensional anisotropic expansion
with shear viscosity.
We show that the dual geometry determines some of the hydrodynamic
quantities in terms of the initial condition and the fundamental
constants. In Section \ref{flow} we show that the hydrodynamics
on the Kasner spacetime
(and hence its gravity dual) describes
an elliptic flow on the flat spacetime
within a well-controlled approximation.
We analyze the properties of the flow.
We conclude in the final section.
The definition of the approximation and its justification are
given in appendix.

\section{Hydrodynamics in Kasner space and anisotropic expansion \label{hydro}}

The remarkable success of Bjorken's hydrodynamics needs to be
extended for more realistic cases of
three-dimensional expansions. The Bjorken's basic assumption is
existence of the so-called ``central rapidity region'' (CPR). The
local rest frame of the fluid can be given by
propertime($\tau$)-rapidity($y$), whose relationship with the
cartesian coordinate is
$(X^{0},X^{1},X^{2},X^{3})=(\tau \cosh y,\tau \sinh y,X^{2},X^{3})$.
We have chosen the collision axis to
be in the $x^{1}$ direction. The Minkowski metric in this coordinate
has the form
\begin{eqnarray}
ds^{2}=-d\tau^{2}+\tau^{2}dy^{2}+(dX^{2})^{2}+(dX^{3})^{2}.
\label{g00}
\end{eqnarray}
Our starting point is an observation that the above metric describes
a one-dimensional Hubble expansion, i.e., an expansion of the
universe instead of that of the fluid, in which the rapidity plays
the role of the co-moving coordinate. The hydrodynamic equations can
be derived from the covariant conservation of the energy-momentum
tensor with the above metric.

The real expansion in RHIC is not one-dimensional but a
three-dimensional one, although the dominant flow is along the
collision axis. The idea is that since the one-dimensional
approximation of RHIC fireball expansion is seen as a Hubble flow, a
three-dimensional Hubble flow may describe the RHIC plasma better.
Let us begin with an ansatz for a local rest frame given by
\begin{eqnarray}
ds^{2}=-d\tau^{2}+\tau^{2a}(dx^{1})^{2}
+\tau^{2b}(dx^{2})^{2}+\tau^{2c}(dx^{3})^{2}.
\label{kasner}
\end{eqnarray}
Here $x^{i},i=1, 2, 3$ denote the co-moving coordinates.
The $a, b$ and $c$ are arbitrary constants for a moment. However, as
we will show in Section \ref{gravity}, they satisfy \be
a+b+c=1,\quad a^2+b^2+c^2=1, \label{bdeq} \ee if we impose conformal
invariance of the fluid. Under the above conditions, the metric is
called Kasner metric describing a homogeneous but anisotropic
expansion of the Universe. In short, we identify the ``Little Bang''
in ``SYM version of RHIC'' with a Big Bang with the homogeneous but
anisotropic expansion described by the Kasner metric.
We use this metric in the late time regime
because we describe only the late time
evolution of the fluid where hydrodynamics is valid.
Therefore the initial
singularity of the metric is not a relevant feature for us. Since we
take $x^1$ as the longitudinal direction, a realistic set up is to
choose
$a\sim 1$ and $b,c \sim 0$ (see section \ref{flow}).

We first establish the hydrodynamics in the Kasner metric.%
\footnote{We use the late time approximation, which is employed in
\cite{sin-shin}, where the macroscopic quantities are assumed to
evolve sufficiently slowly. } We assume that the expansion is
anisotropic but homogeneous, and the physical quantities depend only
on $\tau$. Using the fact that the energy-momentum tensor is
diagonal on the local rest frame, and using the symmetry in transverse
coordinates, we can write \be T^{\mu}_{~~\nu}={\rm
diag}(-\rho,f_1,f_2,f_3), \label{general} \ee where $\rho$ is the
energy density of the fluid. By use of the conservation law $
\nabla_{\mu}T^{\mu\nu}=0$ we get \be {\dot \rho}+\rho/\tau+\sum_i
a_if_i/\tau=0,\label{consv} \ee and from the conformal invariance
$T^{\mu}_{\mu}=0$, we get \be -\rho+\sum f_i=0. \label{cfinv} \ee To
get the above results, we have used the following non-zero
$\Gamma^{\mu}_{\alpha\beta}$'s: \be \Gamma^{\tau}_{11}=a\tau^{2a-1},
\quad \Gamma^{\tau}_{22}=b\tau^{2b-1},\quad
\Gamma^{\tau}_{33}=c\tau^{2c-1} , \quad \Gamma^{1}_{1\tau}=a/\tau,
\quad \Gamma^{2}_{2\tau}=b/\tau, \quad \Gamma^{1}_{3\tau}=c/\tau.
\ee

On the other hand, the energy-momentum tensor in the framework of relativistic hydrodynamics is known to be
\begin{eqnarray}
T^{\mu\nu}=(\rho+P)u^{\mu}u^{\nu}+Pg^{\mu\nu}+\tau^{\mu\nu},
\label{T-ideal}
\end{eqnarray}
where $P$ is the pressure of the fluid, $u^{\mu}=(\gamma, \gamma
\vec{v})$ is the four-velocity field in terms of the local fluid
velocity $\vec{v}$, and  $\tau^{\mu\nu}$ is the dissipative term. In
a frame where the energy three-flux vanishes,
$\tau^{\mu\nu}$ is given in terms of the bulk viscosity
$\xi$ and the shear viscosity $\eta$ by
\begin{eqnarray}
\tau^{\mu\nu}
=-\eta
(\bigtriangleup^{\mu\lambda}\nabla_{\lambda}u^{\nu}
+\bigtriangleup^{\nu\lambda}\nabla_{\lambda}u^{\mu}
-\frac{2}{3}\bigtriangleup^{\mu\nu}\nabla_{\lambda}u^{\lambda})
-\xi \bigtriangleup^{\mu\nu}\nabla_{\lambda}u^{\lambda},
\label{T-dissp}
\end{eqnarray}
under the assumption that $\tau^{\mu\nu}$ is of first order in
gradients. We have defined the three-frame projection tensor as
$\bigtriangleup^{\mu\nu}=g^{\mu\nu}+u^{\mu}u^{\nu}$.
For the conformal invariance, we set $\xi=0$.
Notice that the bulk viscosity in the realistic RHIC setup is
also negligible. (See for example, Ref. \cite{bulk-zero}.)

The four-velocity of the fluid at any point in  the local
rest frame is
$u^{\mu}=(1,0,0,0)$, and this makes the energy-momentum tensor
diagonal. Using $\Delta^{\mu\nu}={\rm
diag}(0,\tau^{-2a},\tau^{-2b},\tau^{-2c})$, $\nabla_\lambda
u^\nu=\Gamma^\nu_{\lambda 0}={\rm diag}(0,a/\tau,b/\tau,c/\tau)$ and
$\nabla_\nu u^\nu =(a+b+c)/\tau$, we get the mixed energy-momentum
tensor:
\begin{eqnarray}
T^{\mu}_{~~\nu}
=
\left(
  \begin{array}{cccc}
  - \rho&   0&   0&   0\\
      0&  P-\frac{2}{3}(3a-1)\frac{\eta}{\tau}
      &   0&   0\\
      0&   0&
  P-\frac{2}{3}(3b-1)\frac{\eta}{\tau} &   0\\
      0&   0&   0&
       P-\frac{2}{3}(3c-1)\frac{\eta}{\tau} \\
  \end{array}
\right). \label{T-diag}
\end{eqnarray}
By identifying (\ref{T-diag}) with (\ref{general}), we obtain
\be
f_i= p-(a_i-{1\over3})\frac{2\eta}{\tau},\ee
where $a_i=a,b,c$ for $i=1,2,3$ respectively.
Inserting these into (\ref{consv}) and (\ref{cfinv}),
we get
\be
{\dot \rho}+{1\over\tau}\rho+
\sum_i{a_i\over\tau}\left(p-(a_i-{1\over 3}){2\eta\over\tau}\right)=0,
\label{consv-2}
\ee
and
\be
-\rho+3p-2\eta\left({{\sum_ia_i-1}\over\tau}\right)=0.
\label{cfinv-2}
\ee
Using the second equation, we may write the first equation in terms
of the energy density as
\be {\dot
\rho}+\left(1+\frac{1}{3}\sum_i{a_i} \right){\rho\over\tau} =
\left(\sum_i a_i^2-\frac{1}{3}(\sum_ia_i)^2\right)
{2\eta\over\tau^2} . \ee
As we will discuss in Section \ref{gravity},
{\it if we impose the conformal invariance}, we get the conditions
\be
\sum_i a_i=1, \;\;\; \sum_ia_i^2=1,
\label{abc-cond}
\ee
under which
the equation of state and the conservation law become: \ba
p&=&{\rho\over 3},\no
\frac{d\rho}{d\tau}+\frac{4}{3}\frac{\rho}{\tau}&=&\frac{4}{3}\frac{\eta
}{\tau^2}. \label{consv2}\ea Remarkably, the equations for fluid
dynamics are completely {\it independent of the parameters $a,b,c$}
in this case. In fact the dynamical law should not depend on the
initial conditions ($a,b$ and $c$) that reflects the initial
collision geometry. In this respect, (\ref{abc-cond})
leads us a satisfactory consequence.

Notice that both of $\rho$ and $\eta$ depend on the proper time
$\tau$ in general.
Let's assume that the shear viscosity evolves by
$\eta= {\eta_{0}}/{\tau^{\beta}},$
where $\eta_{0}$ is a positive constant.
The solution of (\ref{consv2}) is then given by
\begin{eqnarray}
\rho(\tau) &=&\frac{\rho_{0}}{\tau^{4/3}}
+\frac{4/3}{1/3-\beta} \frac{\eta_{0}}{\tau^{1+\beta}}
 \:\:\:\:({\rm for}\:\: \beta\neq 1/3),
 \label{solution}
 \\ \nonumber
 &=&\frac{\rho_{0}}{\tau^{4/3}}
+\frac{4\eta_{0}}{3} \frac{\ln\tau}{\tau^{4/3}}
\:\:\:\:({\rm for}\:\: \beta=1/3),
\end{eqnarray}
where $\rho_{0}$ is a positive constant.
For $\beta \leq 1/3$ case,
the viscous corrections in the hydrodynamic quantities become
dominant in the late time, which invalidates the hydrodynamic
description. If $\beta > 1/3$, the shear viscosity term
is subleading in the late time as we expect.
Therefore we will consider only the latter case from now on.

The proper-time dependence of the temperature $T$
can be read off by assuming
 the Stefan-Boltzmann's law $\rho \propto T^{4}$:
\begin{eqnarray}
T= T_{0}\left( \frac{1}{\tau^{1/3}} +
\frac{1/3}{1/3-\beta}\frac{\eta_{0}}{\rho_{0}}
\frac{1}{\tau^{\beta}} +\cdots \right).
\label{T-general}
\end{eqnarray}
In the {\em static} finite temperature system of strongly coupled
${\cal N}=4$ SYM theory, it is known that $\eta \propto T^{3}$
\cite{PSS-1}. Let us assume that the result is valid
 in the slowly varying non-static cases.
Then we can set $\beta=1$ hence
\begin{eqnarray}
 \eta=\frac{\eta_{0}}{\tau}.
\label{beta1}
\end{eqnarray}
We know $\rho\sim T^4$ and
$\eta\sim T^{3}$ can   be consistent only if there is an additional
term in (\ref{beta1}), but the correction term is negligible in our
case.
The temperature behavior is then given by
\begin{eqnarray}
T=T_{0}\left( \frac{1}{\tau^{1/3}} -
\frac{1}{2}\frac{\eta_{0}}{\rho_{0}} \frac{1}{\tau}
+\cdots \right).
\end{eqnarray}

We can evaluate the entropy change in the presence of shear
viscosity by using hydrodynamics.
The conservation of energy-momentum tensor can be rewritten as
\begin{eqnarray}
 \frac{d(\sqrt{g}\rho)}{d\tau}+ \frac{d\sqrt{g}}{d\tau}P =
 \frac{4}{3}\frac{\sqrt{g}\eta}{\tau^2}
, \label{3deom}
\end{eqnarray}
where $\sqrt{g}=\tau$ is the volume element in the co-moving coordinate.
By integrating over the unit volume in the co-moving coordinate,
and using the thermodynamic relation between the entropy and energy-work,
\be
dE+PdV=TdS,
\ee
(\ref{3deom}) can be written as
\begin{eqnarray}
T\frac{d(\sqrt{g} s)}{d\tau}=\frac{4}{3}\frac{
\eta\sqrt{g}}{\tau^2},
\end{eqnarray}
where $s$ denotes the entropy density and $\sqrt{g} s \equiv S$
is the entropy per unit  co-moving volume.
Notice that in the absence of viscosity, the entropy per unit
co-moving volume is constant.
Now, the entropy per unit co-moving volume has the time dependence:
\begin{eqnarray}
S(\tau) &=& S_0+ \frac{4}{3} \int_0^\tau d\tau \frac{ \eta\sqrt{g}}{
\tau^{2} T}
\nonumber \\
&=& S_\infty - 2\frac{\eta_{0}}{T_{0}}\tau^{-2/3} + \cdots,
\label{entropy}
\end{eqnarray}
where $S_\infty = S(\infty)$. The dissipation creates entropy but
its rate slows down with time. Notice that all these arguments are
completely in parallel with the case of the one-dimensional
expansion.

The hydrodynamics does not calculate the constants $S_\infty$ and
$T_0$ in terms of the initial condition. It is  important to  point
out that by embedding the hydrodynamics into AdS/CFT, we can
determine these parameters in terms of the initial condition
$\rho_{0}$ and the fundamental constants of the theory.

\section{Holographic dual of anisotropic expansion }
\label{gravity}

In this section, we will find a five-dimensional metric that is dual
to the hydrodynamic description of the YM fluid in the previous
section. Some of the parameters of hydrodynamics will be determined
as a consequence. The basic strategy is to use the Einstein's
equation together with the boundary condition given by the
four-dimensional energy-momentum tensor \cite{HSS,Skend,Janik}.
We consider general asymptotically AdS metrics in the
Fefferman-Graham coordinate:
\begin{eqnarray}
ds^{2}=
r_{0}^{2}
\frac{g_{\mu\nu}dx^{\mu}dx^{\nu}+dz^{2}}{z^{2}},
\label{FG-metric}
\end{eqnarray}
where $x^{\mu}=(\tau,x^1,x^{2},x^{3})$ in our case.
$r_{0}\equiv (4\pi g_{s} N_{c} \alpha'^{2})^{1/4}$
is the length scale given by the string coupling $g_{s}$ and
the number of the colors $N_{c}$.
The four-dimensional metric $g_{\mu\nu}$ is expanded
with respect to $z$ in the following form \cite{HSS,Skend}:
\begin{eqnarray}
g_{\mu\nu}(\tau, z)=
g^{(0)}_{\mu\nu}(\tau)+z^{2}g^{(2)}_{\mu\nu}(\tau)
+z^{4}g^{(4)}_{\mu\nu}(\tau)
+z^{6}g^{(6)}_{\mu\nu}(\tau)+\cdots .
\label{expansion}
\end{eqnarray}
Here $g^{(0)}_{\mu\nu}$ is the physical four-dimensional metric for
the gauge theory on the boundary, which is given by (\ref{kasner}) in
the present case. The $g_{\mu\nu}^{(n)}$'s depend only on $\tau$
because our physical quantities are assumed to depend only on
$\tau$.

$g^{(2)}_{\mu\nu}$ is related to the conformal anomaly of the
YM theory in the following way \cite{HSS}:
\begin{eqnarray}
\langle T^{\mu}_{\mu} \rangle
=-\frac{1}{16\pi G_{5}}
\left[
({\rm Tr}g^{(2)})^{2}-{\rm Tr}(g^{(2)})^{2}
\right],
\label{anomaly}
\end{eqnarray}
where $G_{5}$ is the 5d Newton's constant given by
$G_{5}=8\pi^{3}\alpha'^{4}g_{s}^{2}/r_{0}^{5}$ so that
\begin{eqnarray}
\frac{4\pi G_{5}}{r_{0}^{3}}=\frac{2\pi^2}{N_c^2}
\end{eqnarray}
in our notation. Since we are dealing with ${\cal N}=4$ SYM theory
on $R^{1,3}$, we should require the conformal invariance: $\langle
T^{\mu}_{\mu} \rangle=0$. The most natural choice is given by
\begin{eqnarray}
g^{(2)}_{\mu\nu}=0.
\label{g2}
\end{eqnarray}
This is equivalent to the Ricci flat condition for the
four-dimensional metric:
\begin{eqnarray}
R_{\mu\nu}=0,
\label{ricciflat}
\end{eqnarray}
through the relationship \cite{HSS}
\begin{eqnarray}
g^{(2)}_{\mu\nu}
=\frac{1}{2}
\left(
R_{\mu\nu}-\frac{1}{6}R g^{(0)}_{\mu\nu}
\right).
\label{g2-g0}
\end{eqnarray}
Since
\be
R_{00}=(\sum_i a_i-\sum_i a_i^2)/\tau^2, \quad \quad
R_{ii}=a_i(\sum_j a_j-1)\tau^{2a_i-2},
\ee
the Ricci flat condition (\ref{ricciflat}) gives
the Kasner condition (\ref{bdeq}).

One should notice that we are {\em not} solving the four-dimensional
Einstein's equation in the presence of $T_{\mu\nu}$. The Kasner
metric is not a consequence of the gravitational effect of
$T_{\mu\nu}$, but an effective description of the spacetime
expansion satisfying the conformal invariance of strong gauge theory
interaction.

We can identify the first non-trivial data in (\ref{expansion}),
$g^{(4)}_{\mu\nu}$, with the energy-momentum tensor at the boundary
\cite{HSS}:
\begin{eqnarray}
g^{(4)}_{\mu\nu}
=\frac{4\pi G_{5}}{r_{0}^{3}} \langle T_{\mu\nu} \rangle,
\end{eqnarray}
in our notation.
For the time being, we set 
$4\pi G_{5}=1$ and $r_{0}=1$.
The higher-order terms in (\ref{expansion}) are determined by
solving the Einstein's equation with the negative cosmological
constant $\Lambda=-6$ \cite{HSS} (see also \cite{Janik}):
\begin{eqnarray}
R_{MN}-\frac{1}{2}G_{MN}R-6G_{MN}=0,
\label{Eeq}
\end{eqnarray}
where the metric and the curvature tensor are the five-dimensional
ones of (\ref{FG-metric}). $g_{\mu\nu}^{(2n)}$ is described by
$g_{\mu\nu}^{(2n-2)}, g_{\mu\nu}^{(2n-4)}, \cdots, g_{\mu\nu}^{(0)}$
through the Einstein's equation in five dimensions. In other words,
we can obtain the higher-order terms in (\ref{expansion})
recursively by starting with the initial data $g_{\mu\nu}^{(0)}$
($\sim$Kasner) and $g_{\mu\nu}^{(4)}$ ($\sim T_{\mu\nu}$).

Let us come back to our main interest  to obtain
the bulk geometry in the presence of shear viscosity.
The energy-momentum tensor for $\beta=1$ is
written by using  (\ref{solution}) as
\begin{eqnarray}
T^{\mu}_{\nu}=\left(
  \begin{array}{cccc}
    -\frac{\rho_{0}}{\tau^{4/3}}
    +\frac{2\eta_{0}}{\tau^{2}}
       & 0  & 0  & 0  \\
     0 &   \frac{\rho_{0}}{3\tau^{4/3}}
    -\frac{2a\eta_{0}}{\tau^{2}}   & 0  & 0  \\
     0 & 0  &  \frac{\rho_{0}}{3\tau^{4/3}}
    -\frac{2b\eta_{0}}{\tau^{2}}  & 0  \\
     0 & 0  & 0  &  \frac{\rho_{0}}{3\tau^{4/3}}
    -\frac{2c\eta_{0}}{\tau^{2}} \\
  \end{array}
\right).
\label{Tij-viscous}
\end{eqnarray}
We find that the metric components,
$g_{\tau\tau}$, $g_{yy}/\tau^{2}$, $g_{xx}$
have the following structure by solving the Einstein's equation
recursively:
\begin{eqnarray}
f^{(1)}(v)+\eta_{0}h^{(1)}(v)/\tau^{2/3}+
{\tilde f}^{(2)}(v)/\tau^{4/3}+\cdots ,
\label{expansion2}
\end{eqnarray}
Note that {\it the viscosity dependent terms exist at the order of
$\tau^{-2/3}$ and these are more important than the higher-order
terms neglected in} \cite{Janik}. We are considering the late time
region $\tau \gg 1$. But to see the effects of the viscosity, we
need to keep the terms at least to the order of $\tau^{-2/3}$. In
this paper we consider the viscosity effects to the minimal order.

Now we solve the Einstein's equation recursively.
The power series that appear in the solution
can be re-summed to give a compact form of the metric.
After some hard work,
we  can get the late time 5d bulk geometry given by%
\footnote{ One should keep in mind that we are
looking for the late time geometry; the metric (\ref{our-geometry})
is correct only to the order of $\gamma$ and the $O(\gamma^2)$
contributions are not unambiguously determined. The representation
of (\ref{our-geometry}) is chosen since it makes the volume of the
horizon finite.}
\begin{eqnarray}
ds^{2}
&=&
\frac{1}{z^{2}}
\left\{
-\frac{(1-\frac{\rho z^{4}}{3})^{2}}{1+\frac{\rho z^{4}}{3}}
d\tau^{2}
+
\left(1+\frac{\rho z^{4}}{3}\right)
\sum_{i=1}^3
\left(\frac{1+\frac{\rho z^{4}}{3}}{1-\frac{\rho z^{4}}{3}}\right)^{(1-3a_i)\gamma}
\tau^{2a_i}(dx^i)^{2}
\right\}+\frac{dz^{2}}{z^{2}},
\label{our-geometry}
\end{eqnarray}
where
\begin{eqnarray}
\gamma \equiv \frac{\eta_{0}}{\rho_{0}\tau^{2/3}}
 ~~{\rm and}~~
\rho=
\frac{\rho_{0}}{\tau^{4/3}}
-
\frac{2\eta_{0}}{\tau^{2}}.
\label{def-gamma-1}
\end{eqnarray}
Notice that the energy-momentum tensor (\ref{Tij-viscous}) can NOT
be written in terms of the whole $\rho(\tau)$. It is an amusing
surprise to see that the final metric nevertheless can be written in
terms of $\rho(\tau)$  (apart from the powers) in a compact form.
This implies that the position of the horizon can be determined
solely by the energy density.

\subsection{Macroscopic Parameters from Gravity Dual }
The Hawking temperature in the adiabatic approximation
is given by
\be
T(\tau)=\sqrt{2}/(\pi z_{0}(\tau)),
\ee
where
\be
z_{0}(\tau)=[3/\rho(\tau)]^{1/4}
\ee
 is the time-dependent position of the horizon.
 Using  these, we obtain
\begin{eqnarray}
\rho=
\frac{3}{8}\pi^{2}N_{c}^{2}T^{4}(\tau),
\end{eqnarray}
by restoring the normalization of $\rho$: $\rho\to 4\pi G_{5}/r_0^3 \cdot \rho$.
The entropy per unit co-moving volume
is given by
\begin{eqnarray}
S
&=& \lim_{z\to z_0(\tau)} \frac{1}{4G_{5}} \sqrt{\prod_{i=1}^3\left(1+\frac{\rho z^{4}}{3}\right)
\left(\frac{1+\frac{\rho z^{4}}{3}}{1-\frac{\rho z^{4}}{3}}\right)^{(1-3a_i)\gamma}
\tau^{2a_i} }\nonumber \\
&=&\frac{1}{4G_{5}} \frac{2\sqrt{2}\tau r_{0}^{3}}{z_{0}^{3}(\tau)}
\nonumber \\
&=&
N_c^2 \frac{\sqrt{2}}{\pi} \left(\frac{2\pi^2}{N_{c}^{2}}\frac{\rho_0}{3}\right)^{3/4}
\left(
1-\frac{3}{2} \frac{\eta_{0}}{\rho_{0}\tau^{2/3}}
+O(\tau^{-4/3})
\right).
\label{s-evolve-ads}
\end{eqnarray}
Notice that {\it the entropy is completely independent of geometric
 parameters (i.e., $a_i$).}  Without conformal invariance this result is not guaranteed.
As we emphasized in \cite{sin-shin}, the value of $S$ at
$\tau=\infty$, which cannot be determined by hydrodynamics alone, is
precisely determined to be
\begin{eqnarray}
S_{\infty}
=N_c^2 \frac{\sqrt{2}}{\pi} \left(\frac{2\pi^2}{N_{c}^{2}}\frac{\rho_0}{3}\right)^{3/4}
\label{Sinfty}
\end{eqnarray}
in terms of the initial condition $\rho_{0}$ and the microscopic
gauge theory parameter $N_c$. Similarly, \be
T(\tau)=\frac{T_0}{\tau^{1/3}}(1-2\gamma(\tau))^{1/4}, \quad {\rm
with} ~~
T_0=\frac{\sqrt{2}}{\pi}\left(\frac{2\pi^2}{N_{c}^{2}}\frac{\rho_0}{3}\right)^{1/4}.
\label{T0} \ee These parameters $S_\infty, T_0$  are precisely the
quantities used in macroscopic theory (hydrodynamics) which should
be provided by a microscopic theory like QCD. What we are showing
here is that by considering the AdS/CFT dual of hydrodynamics, one
can determine such quantities.

Let us check consistency of (\ref{s-evolve-ads}) and (\ref{entropy}).
Apparent time dependence agrees in the leading order.
In fact one can do more.
The normalized entropy-creation rate from the gravity dual result (\ref{s-evolve-ads})
 is given by
\begin{eqnarray}
\frac{1}{S}\frac{dS}{d\tau}
=\frac{\eta_{0}}{\rho_{0}\tau^{5/3}}+O(\tau^{-7/3})
\label{rate-ads},
\end{eqnarray}
 and that from hydrodynamics result (\ref{entropy}) is
\begin{eqnarray}
\frac{1}{S}\frac{dS}{d\tau}
=\frac{4}{3}\frac{\eta_{0}}{T_{0}S_{\infty}\tau^{5/3}}
+O(\tau^{-7/3}).
\label{rate-hydro}
\end{eqnarray}
Comparing
(\ref{rate-ads}) and (\ref{rate-hydro}), we obtain
\begin{eqnarray}
S_{\infty}=\frac{4}{3}\frac{\rho_{0}}{T_{0}}
=
\left. \frac{4}{3}\frac{\rho \tau}{T}\right|_{\tau=\infty}. \label{consistency}
\end{eqnarray}
This is  the  consistency condition that is required to be checked.%
\footnote{In fact this is the relationship among the entropy, the
energy (per unit co-moving volume) and the temperature obtained by
thermodynamics at $\tau=\infty$ where the system reaches thermal
equilibrium.} With  use of $S_\infty$ and $T_0$ given in eqn's
(\ref{Sinfty}) and (\ref{T0}) we can check that the consistency
condition (\ref{consistency}) is indeed satisfied.

\section{Flow of RHIC fireball and Kasner spacetime \label{flow}}

So far, we have established the gravity dual of
the Yang-Mills system in Kasner spacetime.
Now we would like to suggest a relevance of
our model to description of the elliptic flow in RHIC experiments.
We have seen in section \ref{hydro} that the hydrodynamic
description of the three-dimensional expansion in the
Kasner spacetime is as simple as that of the Bjorken expansion
in the flat spacetime.
Therefore, it is great if we can apply such a simple formalism
to anisotropically expanding RHIC fireball.
However, from  the realistic point of view, we sacrificed the flatness of
4d spacetime for the simplicity of the fluid dynamics of 3 dimensional
expansion.
One immediate question is when and under which condition we can justify it.
The hydrodynamics
on the Kasner spacetime provides a well-approximated
 description of a three-dimensional expansion in the flat
spacetime  if curvature
which is small enough. Notice also that
the spacetime symmetry crucial to our problem, which
is uniformity of the spacetime, is maintained in Kasner spacetime.

\begin{figure}
\centerline{\epsfig{file=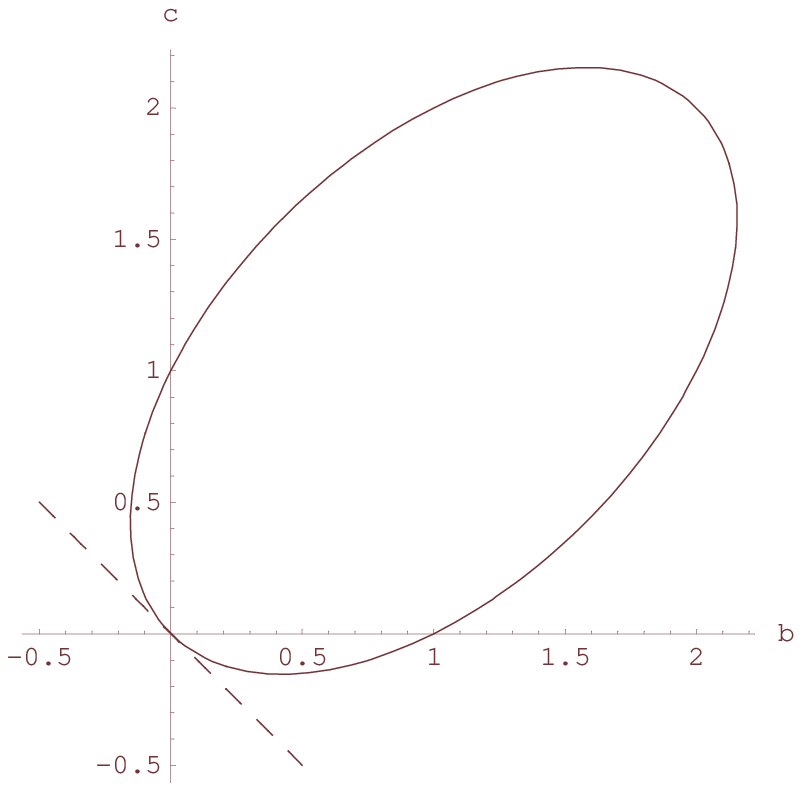,width=6.0cm}}
\caption{\small The available region of $(b,c)$ is on the ellipse
between $(0,0)$ and $(1,0)$. $(b,c)=(0,0)$ is   Bjorken point}
\label{sus}
\end{figure}

In figure 1, we show the allowed region of the anisotropic
parameters.
$(b,c)=(0,0)$ corresponds to the Bjorken expansion.
The non-zero components of the Riemann tensor of the Kasner
spacetime are
\be
R_{0i0i}=(1-a_i)a_i \tau^{2a_i-2}, \quad
R_{ijij}=a_ia_j\tau^{2a_i+2a_j-2},
\ee
and the non-flatness is directly related to the distance
from the Bjorken point $(a,b,c)=(1,0,0)$ on the parameter space.
In order to keep the deviation from the flat spacetime small,
we restrict ourselves within the vicinity of the Bjorken point,
\be
a\simeq 1, \quad b\simeq 0, \quad c\simeq 0,
\ee
that corresponds to almost central collisions in RHIC.

In appendix, we show
 what approximation is necessary to reach kasner space
 starting from a flat spacetime.
The conditions are:
1) the fluid is produced by almost central collision (small $b$),
2) we consider only  the central part of the fluid (small $x^\perp$), and
3) we consider only  the late time regime.
 For more detail, see appendix \ref{appendix}.

With these limitations in mind,
let us consider how the elliptic flow can be described
within our framework.
We can choose $b \ge c$ without any loss of generality.
By considering the intersection of the plane and the unit
sphere in the $a, b, c$ space
of (\ref{bdeq}), we can see that
\be
a>b>-c>0.
\ee
This means that one of the transverse directions must
{\it contract} and the others
expand so that the expansion is elliptical whose eccentricity
grows and eventually saturate to 1.
\footnote{
There is no contraction in the realistic RHIC QGP.
The contraction in the present model is due to the conformal
invariance which is unavoidable for ${\cal N}=4$ SYM theory.
Notice that this volume-preserving nature indicates our
${\cal N}=4$ SYM plasma is more ``liquid-like'' than the
RHIC QGP.
}

The transverse expansion has a natural interpretation as
the elliptic flow, one of the most concrete evidence for the
strong nature of the interaction.
Let $\varepsilon$ be the eccentricity defined by%
\footnote{ The $\varepsilon$ defined here contains an extra minus
sign comparing to other literature such as \cite{KH,bulk-zero}. We
define its positivity by the direction of the $v_2$ evolution. }
\be
\varepsilon=\frac{\langle X_2^2-X_3^2 \rangle_{X}} {\langle
X_2^2+X_3^2 \rangle_{X}},
\label{epsion-def}
\ee where $X^{i}$ is
the cartesian coordinate and
$\langle\cdots\rangle_{X} \equiv \int \cdots \rho \:dX_{2}dX_{3}$.
On the other hand, $v_2$, the quantity experimentally characterizing
the elliptic flow, is
 defined by (see for example, \cite{KH,bulk-zero,Zhang})
\begin{eqnarray}
\frac{1}{N}\frac{dN}{d\phi}
=
\frac{1}{2\pi}
\left(
v_{0}+2v_{2}\cos(2\phi)+2v_{4}\cos(4\phi)+\cdots
\right),
\end{eqnarray}
where $N$ is the number of the partons and $\phi$ is the angular
coordinate on the transverse momentum plane. It can be calculated
from the following identification,
\begin{eqnarray}
v_2=
\frac{\int d^{2}P_{T}
\left(\frac{P_2^2-P_3^2}{P_2^2+P_3^2 }\right)
\frac{dN}{d^{2}P_{T}}}{\int d^{2}P_{T}\frac{d N}{d^{2}P_{T}}},\label{v2def}
\end{eqnarray}
where $P^{i}$ is the momentum of the fluid in the $X^{i}$ coordinate
and $d^{2}P_{T}=dP^{2}dP^{3}$.
To consider $v_2$ in the present model, we introduce a
coarse-grained (i.e., averaged over a small volume)
momentum flow,
\be P^i=K^i(x,\tau). \ee At each fixed time $\tau$, this can be
considered as a mapping from the $P^i$-space to the
$x^i$-space.\footnote{ In the co-moving coordinate, where the fluid is
at rest, we do not have any flow. So $P^i(x)$ is the coarse-grained
momentum field in Minkowski space written as a function of co-moving
coordinate. } Now $v_2$ can be expressed as an integral over the
co-moving coordinates:
\be {  v}_2= \frac{\int dx^{2}dx^{3}
\left(\frac{P_2^2-P_3^2}{P_2^2+P_3^2 }\right)_x \rho_N(x)}{\int
dx^{2}dx^{3}\rho_N(x)},
\label{v2-comp}
\ee
where $\rho_N(x)=\frac{dN}{dx^2 dx^3}$ is the particle density
in the transverse space.
Notice that the Jacobians cancel out.

To proceed, we need an explicit expression for $K(x)$. We first
relate the coordinates of the Kasner spacetime and the usual
Minkowski spacetime.
We can identify the flat space variable $X^i$ for $i=2, 3$ by
%
\begin{eqnarray}
X^i \simeq \tau^{a_i}x^i ~~{\rm near}
~~x^i=0,
\label{x-trans}
\end{eqnarray}
which is nothing but (\ref{transform}) under (\ref{approximation}).
Then the fluid momentum in the flat space is given by \be
P^{i} =\rho  {dX^i\over d\tau}=\rho  \:a_i\tau^{a_i-1} x^{i}. \ee
Here $\rho =m\rho_N$ is a mass density with some proper mass
parameter $m$. Since we treat it as a constant from now on, it is
irrelevant in the calculation below.

A few technical remarks are in order:
\begin{enumerate}
 \item One should notice that the fluid momentum
$P^i(x)$ is different from the individual particle momentum. By
replacing the momentum by fluid momentum, we expect a small
deviation from the original $v_2$. So one may want to call the final
expression by $\bar  v_2$. However, this is precisely what we should
have when we describe the system by hydrodynamics where everything
is to be defined in the coarse-grained level.
  \item Since our approximation is valid in the small $x^{i}$ region,
the integrals in (\ref{v2-comp}) are now defined within
the small $x^{i}$ region by introducing a cut off radius.
\end{enumerate}

The resulting $v_2$ can be calculated in our model to yield
\be {
v}_2=\frac{b\tau^{b}+ c\tau^{c} }{b\tau^{b}- c\tau^{c}},
\label{v2}
\ee
for $b>0, c<0$. Comparing this with the result for the
eccentricity in the small $x^{i}$ region
\be \varepsilon=\frac{
\tau^{2b} -\tau^{2c} }{ \tau^{2b}+ \tau^{2c}},
\label{ecc}
\ee
we plot the time evolution of $v_2$ in figure 2.

Notice that the hydrodynamics describes relatively late time
regime and we do not consider the time region where $v_2$ is
negative.
At the time of  zero $v_2$, $\varepsilon $ is negative
and its absolute value decreases
to zero. That is, the fireball core becomes more round in the
transverse space.
After that initial period, $\varepsilon$ becomes positive
and grows in the same direction of growing $v_2$.
This is qualitatively the same behaviour as those given in
\cite{KH,bulk-zero} and \cite{Zhang}.
Notice that there is only one independent parameter $b$
that can be used to parameterize the initial eccentricity.

\begin{figure}
\centerline{\epsfig{file=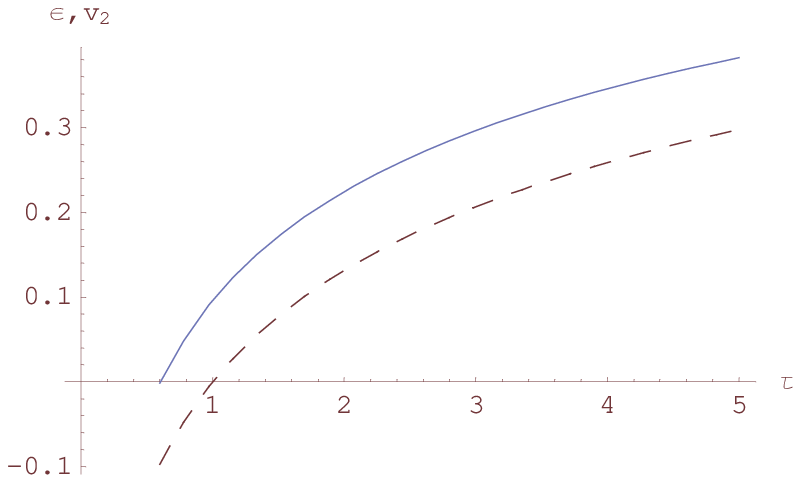,width=6.0cm}}
\caption{\small Time evolution of the elliptic flow parameters.
Dotted line is the eccentricity and the solid line is the $v_2$ for
$b=0.1$.}
\label{sus}
\end{figure}

\section{Discussion}
In this paper, we extended the Bjorken hydrodynamics to the case of
anisotropic three-dimensional expansion. The Ricci flat condition
suggested by the conformal invariance imposes the condition on the
anisotropy parameters. As a consequence, the four-dimensional
boundary metric becomes precisely that of Kasner universe.
Our hydrodynamic equation of motion is independent
of the anisotropic parameters, so that it is the
same as that obtained by Bjorken.

Although the Kasner spacetime is a curved spacetime, we found
that it gives a well-approximated local rest frame
of an anisotropically expanding fluid on the flat spacetime.
We obtained the eccentricity and $v_{2}$ of the elliptic flow
of the fluid.
The expansion in this setting
has a deviation from that of realistic RHIC fireballs,
since one of the
transverse direction contracts as a consequence of the
Kasner condition. This deviation is essentially due to
the conformal invariance. Whether one can lessen the condition
of the conformal invariance is an important issue,
which we will treat in other publication.

We also extended the ``falling horizon solution'' obtained in
\cite{Janik,sin-shin} to the case of three-dimensionally expanding
fluid. It is very important to figure out how to use the bulk metric
apart from reading off the horizon location, which gives the cooling
rate and the entropy creation rate. It is also interesting to
workout the details of the Hawking evaporation and Wilson line
calculations for the external quark-antiquark in our time-dependent
metric. We will come back to these issues in later publication.


\noindent
{\large\bf Acknowledgments}\\
We thank T. Hirano and E. Shuryak for valuable comments
 on the manuscript and interesting suggestions.
S.N. thanks the Yukawa Institute for Theoretical Physics at Kyoto University.
Discussions during the YITP workshop YITP-W-99-99 on
``thermal quantum field theories and their applications''
were useful to improve the section 4.
This work was supported by
the SRC Program of the KOSEF through the Center for Quantum
Space-time(CQUeST) of Sogang University with grant number R11 - 2005
-021 and also by KOSEF Grant R01-2004-000-10520-0.

\appendix
\section{Relationship between Kasner spacetime and our
local rest frame}
\label{appendix}

We clarify the relationship between our local rest
frame and the Kasner spacetime in the following strategy.
We start from the cartesian coordinate and make a coordinate
transformation to define our local rest frame. The transformation
is essentially what we have employed at (\ref{x-trans}).
At this stage, we are still on the flat spacetime but
the local rest frame is that of the anisotropically
expanding fluid with elliptic flow.
Next, we see that the local rest frame is equivalent to
the Kasner spacetime under a well-controled approximation
which is specified at the end of this section.

Let us start from the cartesian coordinate,
\begin{eqnarray}
ds^{2}=-(dX^{0})^{2}+(dX^{1})^{2}+(dX^{2})^{2}+(dX^{3})^{2},
\end{eqnarray}
and perform the following coordinate transformation
\begin{eqnarray}
(X^{0},X^{1},X^{2},X^{3})
=(\tau \cosh x^{1},\tau \sinh x^{1},\tau^{b}x^{2},\tau^{c}x^{3})
\label{transform}
\end{eqnarray}
to define our local rest frame $(\tau,x^{1},x^{2},x^{3})$.
Notice that the $(X^{0},X^{1})$ part has been ``boosted'' with
rapidity $x^{1}$ and the
$(X^{2},X^{3})$ part has been transformed by (\ref{x-trans}).
The resulting metric on the local rest frame is given by
\begin{eqnarray}
ds^{2}=
-d\tau^{2}
+\tau^{2a}(\tau^{2-2a})(dx^{1})^{2}
+\tau^{2b}\left(dx^{2}+b\frac{x^{2}}{\tau}d\tau\right)^{2}
+\tau^{2c}\left(dx^{3}+c\frac{x^{3}}{\tau}d\tau\right)^{2}.
\label{lrf-met}
\end{eqnarray}
The parameters $a$, $b$, $c$ are free so far.

Next,
let us impose the Kasner condition (\ref{bdeq}) among $a$, $b$,
$c$ and assume $b\ll 1$.
$a$ and $c$ is then given
by $a=1-3b^{2}$ and $c=-b+3b^{2}$ approximately.
Now (\ref{lrf-met}) is written as
\begin{eqnarray}
ds^{2}&=&
-d\tau^{2}
+\tau^{2a}\left(1+6b^{2}\log\tau+\cdots\right)(dx^{1})^{2}
\nonumber \\
&&+\tau^{2b}\left(dx^{2}+b\frac{x^{2}}{\tau}d\tau\right)^{2}
+\tau^{2c}
\left(dx^{3}+(-b+3b^{2}+\cdots)\frac{x^{3}}{\tau}d\tau\right)^{2}.
\label{lrf-met-2}
\end{eqnarray}
The above metric agrees with the Kasner metric (\ref{kasner}) under
the following approximation:
\begin{eqnarray}
\frac{|x^{2}|}{\tau}\sim \frac{|x^{3}|}{\tau}\sim
b\ll 1, \:\:{\rm with}\:\: b\log\tau\sim O(1).
\label{approximation}
\end{eqnarray}
We ignore $O(b^{2})$ contributions while we keep $ b\log\tau$.
This is a mixture of
small $b$ approximation (we consider almost central collision),
small $(x^{2}, x^{3})$ approximation
(we consider only within the central part of the QGP)
and the late time approximation.
The condition $b\log\tau\sim O(1)$ is necessary for us
to get the nontrivial results.

\end{document}